\documentstyle[twocolumn,epsfig,psfig,aps,prl,amssymb]{revtex}  
\begin{document}

\wideabs{

\title{Oscillating Fracture in Rubber}
\author{ Robert D. Deegan, Paul J. Petersan, M. Marder, and Harry L. Swinney} 
\address{Center for Nonlinear Dynamics and Department of Physics\\ The
  University of Texas at Austin, 
Austin, TX 78712}
\date{\today}
\maketitle 
\begin{abstract}
We have found an oscillating instability of fast-running cracks in thin rubber sheets.  A well-defined transition from straight to oscillating cracks occurs as the amount of biaxial strain increases. Measurements of the amplitude and wavelength of the oscillation near the onset of this instability indicate that the instability is a Hopf bifurcation.
\end{abstract}  
}
\draft
 
\narrowtext 
When a balloon is pierced, a crack races around the circumference of the balloon, slicing the material into fragments.   Surprisingly, instead of running straight the crack wiggles, leaving a wavy pattern on the fractured edges of the fragments.  This effect has likely been observed many times, but we found only one documented case~\cite{Stevenson.79}.  We have constructed an experiment to study this phenomenon.    

Elastomers fill an important technological need, and understanding their properties is of commensurate importance.  Aside from practical concerns, the oscillations of cracks in rubber may shed light on the broader issue of crack path selection.   The field of dynamic fracture was spawned by theoretical attempts to understand the onset of crack branching in brittle materials~\cite{Yoffe.51}, an instability marked by the bifurcation of a straight-running crack. Despite the great advances in the understanding of dynamic fracture since then~\cite{freund}, the crack path selection problem remains unsolved.  The oscillation of cracks in rubber may provide a new avenue to address this issue, as this is the first case we know where a rapidly moving crack spontaneously chooses a wavy path in a homogeneous setting. 

There is a vast literature on the fracture of elastomers dating back a half century~\cite{Rivlin.53}.  The applicability of the principles of fracture mechanics to elastomers has been established~\cite{Thomas.94,Lake.95,Extrand.91}, the time-dependent characteristics of elastomer fracture have been examined~\cite{Kadir.81,Lake.91}, and crack speeds under varying degrees of biaxial strain have been investigated~\cite{Gent.82b,Lake.00}.  None of these studies, however, has explored the transition from a straight to a wavy crack path.  Our aim here is to characterize this instability.

\begin{figure}
\begin{center}
\epsfig{file=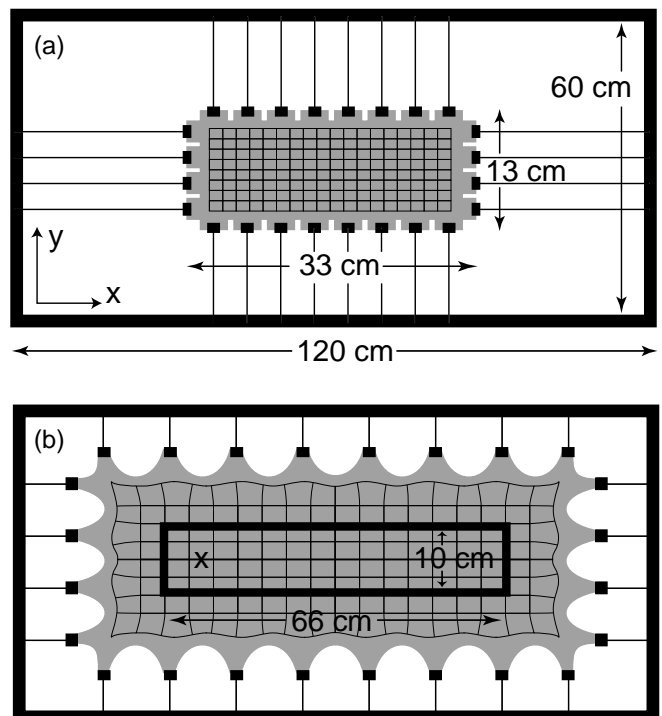, width=\linewidth}
\caption{The experimental apparatus for straining rubber sheets along two axes. (a) A grid is drawn on the sample and clamps are attached to precut tabs along the sample's edges.  The load is applied to the sample through the clamps, which are attached by wires to the rigid outer frame.  (b) After the sheet has been slowly extended in the $x$- and $y$-directions, it is clamped by an inner rectangular frame.  The sheet is distorted near its edges but not inside the inner frame.  After clamping the sheet, it is punctured with a pin at the point marked $\times$.  Since the edges of the sheet are clamped, no energy flows into it during fracture.}
\label{fig:apparatus}
\end{center}
\end{figure}

Inflating a balloon introduces biaxial strain into the material and a pressure drop across the surface.  The pressure difference plays no role so it is sufficient to experiment with flat sheets in biaxial tension.  Our apparatus, inspired by a similar technique developed by Treloar~\cite{Treloar.75}, is shown Fig.~\ref{fig:apparatus}. Most of our rubber samples were taken from a single roll of 0.18~mm thick natural latex sheet.  Tabs were prepared on 32.5~cm~$\times$~12.7~cm sheets by cutting 1.2~cm long slots, perpendicular to the edge, 2.5~cm apart (Fig.~\ref{fig:apparatus}(a)).  The tips of these incisions were rounded by melting them with a soldering iron to prevent cracks from initiating at these points during loading of the sample. 
 
The experiment proceeds by gripping the tabs and incrementally increasing the load simultaneously in both $x$- and $y$-directions, until the desired strain level is reached. The applied strain is on the order of 200\%, uniform within 5\%, and always chosen so that the strain in the $y$-direction, $\epsilon_y$, is greater than the strain in the $x$-direction, $\epsilon_x$.  Strain is measured from the dilation of the grid; deviations from uniform strain are identified from the distortion of the grid and minimized by individually adjusting each clamp.  Once the desired strain level is attained, the rubber sheet is sandwiched between a pair of 10~cm~$\times$~66~cm rectangular steel frames (Fig.~\ref{fig:apparatus}(b)).  The loading is then maintained entirely by the frames.

\begin{figure}[t]
\begin{center}
\epsfig{file=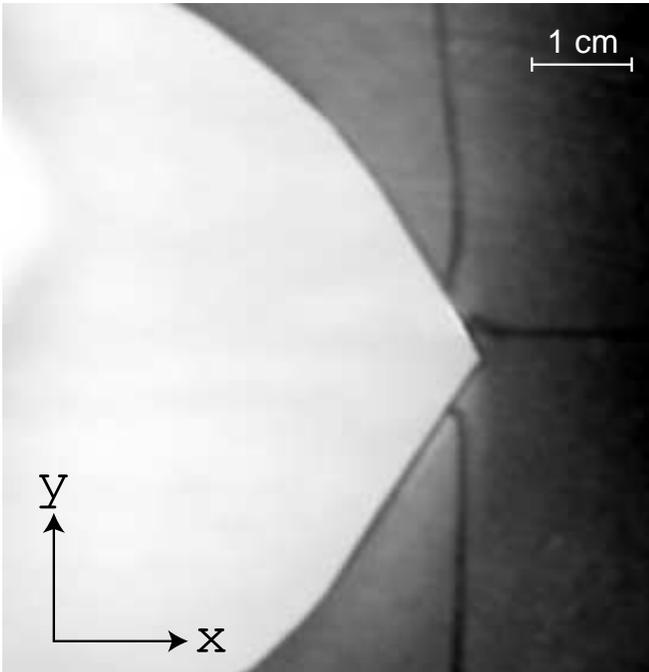, width=\linewidth}
\caption{Snapshot of the tip of a propagating crack.  The crack is moving from left to right.  The initial conditions were $\epsilon_y=2.3$ and $\epsilon_x=1.8$.}
\label{fig:cusp}
\end{center}
\end{figure}

Each run is initiated by pricking the sheet with a pin at the point marked $\times$ in Fig.~\ref{fig:apparatus}(b).  The crack tip that forms is sharp and wedge-shaped, as shown in Fig.~\ref{fig:cusp}.  The crack travels down the centerline, the midpoint between upper and lower edge of the frame.  Depending on the initial strain conditions, the crack runs straight or oscillates about the centerline.  The inner frame makes steady states possible because the energy stored per unit length inside the framed sample is constant ahead of the crack and also constant in its wake.  So as the crack tip advances it consumes a fixed amount of energy per unit length of advancement.  Indeed, we observe that after an initial transient period, an oscillating crack propagates with a wavelength and amplitude that are constant to within $10\%$.

\begin{figure}[h]
\begin{center}
\epsfig{file=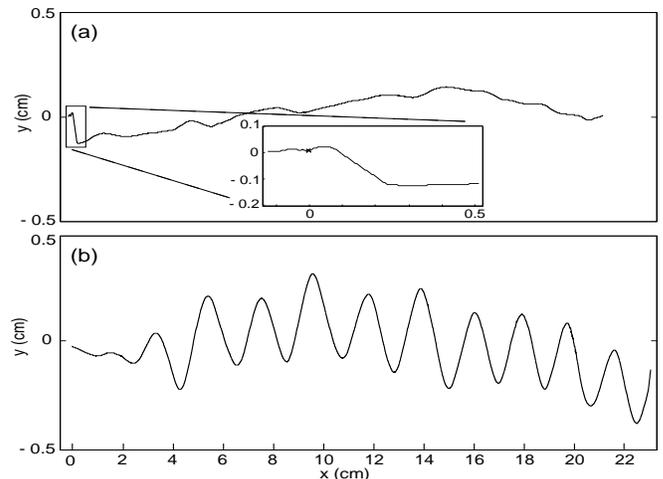, width=\linewidth,height=2.5in}
\caption{(a) A straight crack with $\epsilon_x = 1.2$, $\epsilon_y = 2.0$ (inset shows the initial kink that is sometimes observed) and (b) an oscillating crack with $\epsilon_x = 1.3$, $\epsilon_y = 1.8$ are shown in the final, unstrained state.  In both cases, the crack initiated at the left edge and propagated to the right.  Both samples were $0.18$~mm thick, held by a 10~cm $\times$ 66~cm frame.  }
\label{fig:profiles}
\end{center}
\end{figure}

Examples of the paths of a straight and a wavy crack are shown in Fig.~\ref{fig:profiles}.  These curves were obtained by scanning the cracked sheet with a flatbed scanner and applying an edge-finding algorithm to the resulting image.  Transients typically dominate the first $15\%$ of  the crack length.  During this transient regime a wavy crack's oscillations grow to saturation and a straight crack recovers from any off-centerline starts, as shown in the inset of Fig.~\ref{fig:profiles}(a). 

\begin{figure}[!h]
\begin{center}
\epsfig{file=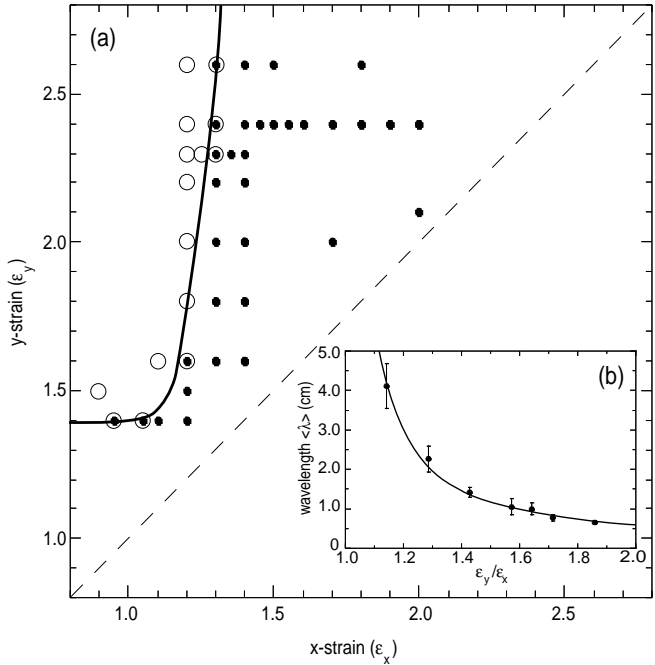, width=\linewidth}
\caption{(a) A phase diagram showing the two observed states:  oscillating cracks $\bullet$, and straight cracks $\circ$, as a function of the initial strain state, $\epsilon_x$ and $\epsilon_y$.  The solid line shows our estimate of the phase boundary between straight and wavy cracks.  Data were not taken in the region below the dashed line because there the principal direction of crack motion is across the width of the frame. (Frame size = 10~cm~$\times$~66~cm) (b) Wavelength versus $\epsilon_y$, normalized by $\epsilon_x$, for fixed $\epsilon_x=1.4$.  The solid line is a fit of the data to $A/(\epsilon_y/\epsilon_x-1)$, which yields $A=0.58$.}
\label{fig:phasediag}
\end{center}
\end{figure}

Rubber sheets with many different initial strain states were fractured, and the crack path was found to undergo a transition from straight to wavy with increasing biaxial strain.  Since the applied forces are purely tensile, the strain state is fully described by $\epsilon_x$ and $\epsilon_y$.
The results of these runs are shown in the phase diagram in Fig.~\ref{fig:phasediag}.  
The control parameter range was limited by experimental difficulties found at the highest and the lowest values of $\epsilon_y$.  For $\epsilon_y>2.6$ it became impossible to complete a run because cracks would spontaneously form at the high stress point between the tabs.  For $\epsilon_y< 1.4$ it became impossible to distinguish between straight and wavy cracks because the wavelength became comparable to the length of the sample, as indicated by the trend in the inset of Fig.~\ref{fig:phasediag}.  Ambiguous points $\odot$ appear in Fig.~\ref{fig:phasediag}(a) at high values of $\epsilon_y$  because different runs with the same initial conditions produced both straight and wavy cracks;  ambiguous points at low values of $\epsilon_y$ result from the difficulty in discriminating between straight and wavy cracks.

\begin{figure}[!h]
\begin{center}
\epsfig{file=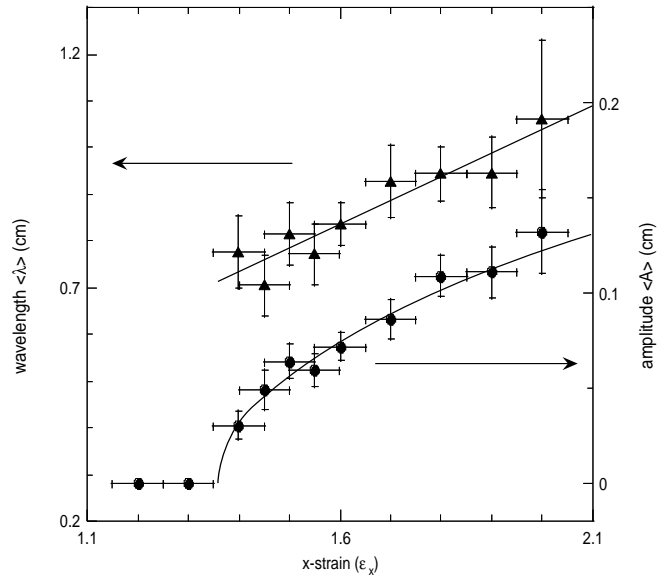, width=\linewidth,height=3in}
\caption{Amplitude $\langle A\rangle$ $\bullet$ and wavelength $\langle\lambda\rangle$ $\triangle$ are plotted across the transition from straight to oscillating cracks.  $\epsilon_y = 2.4$ (see Fig.~\ref{fig:phasediag}).  The solid line through the amplitude data is a square-root fit.}
\label{fig:transition}
\end{center}
\end{figure}

We measured the average wavelength $\langle\lambda\rangle$ and the average amplitude $\langle A\rangle$ of the wavy edge while holding $\epsilon_y=2.4$ fixed and varying $\epsilon_x$ from 1.2 to 2.0.   From the digitized curves we extracted the wavelength as the peak-to-peak distance in the $x$-direction and the amplitude as half the peak-to-valley distance in the $y$-direction, and averaged these quantities over multiple runs with identical initial conditions.  These data are plotted in Fig.~\ref{fig:transition}.

The amplitude grows as the square root of the control parameter with a critical value of $x$-strain, $\epsilon_x=1.36$.   Furthermore, if we assume that the crack travels with a constant velocity $v_c$ in the $x$-direction, then the wavelength is equal to $2\pi v_c/\omega$, where $\omega$ is the frequency at which the crack tip oscillates in the $y$-direction.  Since at onset of the instability the wavelength is nonzero,  the frequency at onset is nonzero. Hence, we conclude that the instability is a Hopf bifurcation.

\begin{figure}[h ]
\begin{center}
\epsfig{file=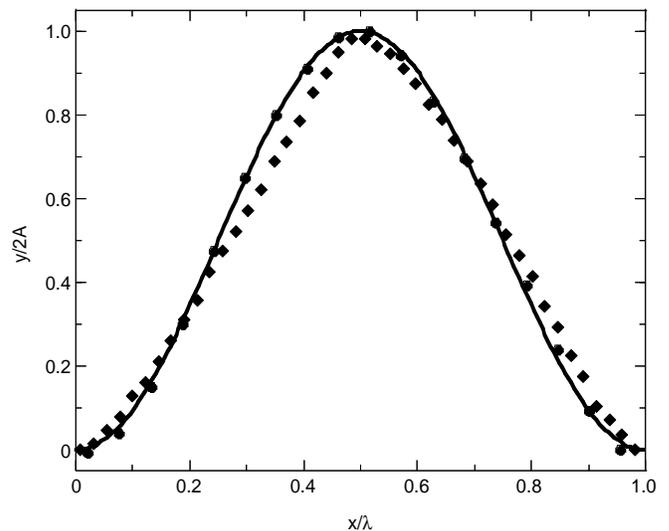, width=\linewidth}
\caption{Profiles of individual peaks from two different runs are scaled by their amplitude and wavelength and plotted together.  Each run was performed under different strain conditions but for the same rubber thickness and frame size.  A sine curve (solid line) is plotted for comparison.($\bullet$:  $\epsilon_x = 1.4$, $\epsilon_y = 2.4$, $A$ = 0.03~cm, $\lambda$ = 0.75~cm; $\blacklozenge$:  $\epsilon_x = 2.0$, $\epsilon_y = 2.4$, $A$ = 0.13~cm, $\lambda$ = 1.02~cm).  }
\label{fig:peaks}
\end{center}
\end{figure}

The waveform of the crack path is nearly sinusoidal over a wide range of amplitudes (0.03~cm to 0.39~cm) and wavelengths (0.56~cm to 5.0~cm).  Fig.~\ref{fig:peaks} illustrates this point with two waveforms taken from the runs used to construct  Fig.~\ref{fig:transition}.  These curves correspond to the first data point after the transition and the last data point in Fig.~\ref{fig:transition}.  The waveform near the transition is almost a perfect sinusoid, consistent with the transition being a Hopf bifurcation.  The waveform far from the transition shows sizeable deviations from a sinusoid and it is skewed in the direction of propagation.

In addition to our quantitative results, we explored the possibility that the oscillation arises from out-of-plane vibrations, strain crystallization, and interaction of the crack tip with waves reflected from the boundary.  In one experiment we reduced out-of-plane motion by sandwiching the rubber sheet snugly between two glass plates after the sheet was stretched to the loaded state.    In another experiment we loaded the sheet as usual and then forced a cylindrical surface into the sheet so that the sheet was everywhere pressed into contact with the surface;  thus the only out-of-plane motion possible is away from the surface.  Neither experiment stopped the crack from oscillating; hence out-of-plane motion is not the source of crack oscillations.  

We also considered the effect of the known propensity of polymer chains in latex rubber to align when strained predominately in one direction~\cite{Treloar.75}.  Measurements of the elastic response of our rubber samples indicate that the initial strains in our experiments were well below the threshold at which strain crystallization and finite chain length effects are significant~\cite{Treloar.75}.  Furthermore, by increasing the strain in both the $x$- and $y$-direction simultaneously so that the strain is never much larger in one direction than the other, we avoided the conditions leading to strain crystallization.  We further considered the possibility that the high stress around the crack tip could locally crystallize the material as the tip passed through.  However, the kinetics of the process is too slow.  Assume that stresses are sufficiently large to crystallize the rubber out to some radius $R$ from the crack tip.  This region is traveling with a velocity $v_c$ and so the maximum time available for crystallization to occur is $R/v_c$.  The value of $R$ must be less than the half-width of the frame, 5~cm, and the lowest crack velocity we measured was 40~m/s.  This places an upper bound of 1.3~ms on time available for the rubber to crystallize.  Comparing this result with the measured nucleation time of 60~ms for rubber at high strains~\cite{Mitchell.68},  we find that there is insufficient time for the material around the crack tip to crystallize.  A similar conclusion was reached by Lake and coworkers after finding similar crack growth rates for both a crystallizing and non-crystallizing natural rubber during high-speed fracture~\cite{Lake.00}.

In brittle materials, waves originating from the crack tip and reflecting from the boundaries are known to produce periodic markings such as those identified as Wallner lines~\cite{Lawn.93}.  If such a mechanism were active in rubber, one would expect an oscillatory wavelength $\lambda=(2h v_c/c)/\sqrt{1-v_c^2/c^2}$ where $h$ is the distance to the boundary.  Given that the vertical boundary is 5~cm from the crack and that our measurements show $v_c/c<0.5$, it follows that $\lambda >  5.7$~cm.   Yet, the smallest wavelength observed was 0.85~cm (in the strained state).  Thus waves reflecting from the boundary cannot account for the oscillations of the crack. 

In conclusion, we have found a well-defined instability in the propagation direction of a crack in rubber sheets with biaxial strain.  This is an unexpected result given the homogenous elastic field in which the crack propagates.  We have shown that this instability can be characterized as a Hopf bifurcation.   We have ruled out strain crystallization, out-of-plane motion and wave reflections from the boundary as possible mechanisms for the oscillation.

We are grateful to Stefan Luding for introducing us to this phenomenon.
We thank Eric Gerde,  Matt Lane, W.D. McCormick,  L. Mahadevan, and Eran Sharon for discussions and suggestions, the National Science Foundation for support under DMR-9802562, and the Vice President for Research at the University of Texas at Austin for an exploratory grant.

\end{document}